\documentclass{article}
\usepackage{graphicx}
\usepackage{subcaption}

\usepackage{xcolor}
\usepackage{amssymb}
\usepackage{float}
\usepackage{amsmath}
\usepackage[a4paper, total={7in, 8in}]{geometry}
\usepackage{soul}
\usepackage{nomencl}
\usepackage[numbers,square]{natbib}
\renewcommand{\nomgroup}[1]{%
  \ifthenelse{\equal{#1}{I}}{\item[\textbf{Indices}]}{%
  \ifthenelse{\equal{#1}{P}}{\item[\textbf{Parameters}]}{%
  \ifthenelse{\equal{#1}{V}}{\item[\textbf{Continuous Variables}]}{%
  \ifthenelse{\equal{#1}{B}}{\item[\textbf{Binary Variables}]}{}}}}}
\makenomenclature
\usepackage{comment}
\usepackage{booktabs}
\usepackage{amsmath}
\usepackage{multirow}
\usepackage{abstract}
\usepackage{url}

\usepackage[hidelinks]{hyperref}
\usepackage{authblk}

\title{Towards Quantum Stochastic Optimization for Energy Systems under Uncertainty: Joint Chance Constraints with Quantum Annealing}

\author{D. Ribes,\enspace T. González Grandón\thanks{Email: tatiana.grandon@ntnu.no}}

\date{\today}

\begin{document}
\maketitle

\begin{abstract}
Uncertainty is fundamental in modern power systems, where renewable generation and fluctuating demand make stochastic optimization indispensable. The chance-constrained unit commitment problem (UCP) captures this uncertainty but rapidly becomes computationally challenging as the number of scenarios grows. Quantum computing has been proposed as a potential route to overcome such scaling barriers. In this work, we evaluate the applicability of quantum annealing platforms to the chance-constrained UCP. Focusing on a scenario approximation, we reformulated the problem as a mixed-integer linear program and solved it using D-Wave’s hybrid quantum–classical solver alongside Gurobi. The hybrid solver proved competitive under strict runtime limits for large scenario sets (15,000 in our experiments), while Gurobi remained superior on smaller cases. QUBO reformulations were also tested, but current annealers cannot accommodate stochastic UCPs due to hardware limits, and deterministic cases suffered from embedding overhead. Our study delineates where chance-constrained UCPs can already be addressed with hybrid quantum–classical methods, and where current quantum annealers remain fundamentally limited.
\end{abstract}


\section{Introduction}

Uncertainty is not an exception in modern systems; it is the rule. From climate instability to volatile energy markets, from digital infrastructures to public health, societies face futures that are not singular and predictable but plural and contingent. For power systems in particular, growing reliance on renewables means that tomorrow’s supply and demand can no longer be forecast as fixed numbers; they must instead be treated as probability distributions \cite{Zakaira2020}. 

Stochastic optimization provides a rigorous way of making decisions in the face of such uncertainty. Yet this rigor comes at a price: the number of possible futures expands exponentially with each new source of randomness. With just ten random variables, each discretized into ten outcomes, the resulting scenario space already contains $10^{10}$ possibilities. Scaling to realistic industrial dimensions only compounds this growth: each additional source of uncertainty multiplies the space of possibilities. This phenomenon, known as the curse of dimensionality, pushes classical computational methods beyond their practical limits \cite{Shapiro2005}.

Quantum computing offers a fundamentally different paradigm. Unlike classical binary bits, which can encode only one configuration out of 
$2^N$ possibilities at a time, qubits can exist in a superposition of all 
$2^N$ basis states simultaneously, each with a probability amplitude that can be manipulated through quantum operations \cite{Grover1996}. 
Quantum entanglement further correlates qubits so that their states cannot be described independently, enabling collective behavior across the system \cite{Einstein1935}. 
This architecture mirrors the probabilistic structure of stochastic optimization: whereas classical methods must enumerate scenarios sequentially, quantum algorithms can exploit superposition and interference to explore many possibilities in parallel. 
Such quantum parallelism positions quantum computing as a natural candidate for addressing the exponential growth of uncertainty. 
Despite this promise, \emph{stochastic quantum optimization remains largely unexplored}.

Among today’s Noisy Intermediate-Scale Quantum (NISQ) devices, gate-based architectures are universal in principle but remain limited by qubit counts and error rates \citep{Lau2022}. Adiabatic Quantum Annealers (QA), such as those developed by D-Wave, are instead tailored to combinatorial optimization \citep{Rajak2023}. Rather than applying a sequence of quantum logic gates, they gradually evolve to a simple and known low-energy state toward the ground state of the optimization problem. This approach is naturally suited to Quadratic Unconstrained Binary Optimization (QUBO) formulations \cite{Quinton2025,Kochenberger2004} because of their correspondence with the Ising Hamiltonian \cite{Lenz1920, Ising1925}.  

Although QAs are advancing rapidly \citep{Dwave_whitepaper}, they are not yet capable of addressing problems at real-world scale. To overcome these limitations, hybrid approaches have emerged as a promising strategy. These combine quantum and classical resources to obtain solutions for complex optimization tasks. Hybrid quantum--classical solvers offer a pragmatic pathway for near-term applications \citep{Dwave2022hybrid}. Examples include Mixed-Integer Quadratic Programs (MIQPs), which involve both binary and continuous decision variables, and Constrained Quadratic Models (CQMs). The latter are particularly advantageous because constraints can be incorporated directly into the model rather than reformulated as tuned penalty terms.

Against this backdrop, the Unit Commitment Problem (UCP) constitutes a natural and practically significant test case \citep{Montero2022}. At the core of short-term power system planning, the UCP prescribes the on/off status and generation levels of thermal units to satisfy demand at minimal cost while meeting operational and security constraints. In modern power systems, rising shares of wind and solar make supply and demand inherently uncertain, so deterministic UCP formulations risk infeasibility and costly reserve use. Chance-constrained formulations address this deficiency by requiring that system constraints hold with high probability \citep{Ouanes2024}. Their computational complexity \citep{Dai2015,Campos2014}, is substantial, making the stochastic UCP with chance constraints an especially appropriate benchmark to assess novel optimization paradigms such as quantum computing.

A small literature has begun to explore quantum approaches for stochastic optimization in power systems. Braun et al. \citep{Braun2023} examined a UCP with uncertain renewable supply, demand, and unit failures using an expected-value formulation, comparing quantum and classical solvers. Rotello et al. \citep{Rotello2024} proposed a two-stage stochastic UCP algorithm that achieves a theoretical time speedup relative to classical methods while also reducing the required number of qubits. Yet, despite these advances, chance-constrained formulations of the UCP have not been addressed on quantum platforms, leaving a significant gap.

In this work, we present, to the best of our knowledge, the first systematic investigation of stochastic optimization with chance constraints on a quantum annealer. Our study proceeds in three steps. First, we formulate the stochastic UCP with joint chance constraints as a Mixed-Integer Linear Program (MILP) and examine its tractability within D-Wave’s hybrid CQM framework under different correlation structures of the stochastic vector. Second, we benchmark the approach in terms of scalability with respect to the number of scenarios, i.e., the growth of the scenario set induced by discretization of multiple random variables, and compare the resulting solution quality against a state-of-the-art classical solver. Finally, we reformulate the UCP as a QUBO problem and introduce a novel algorithm for tuning the penalty factors associated with penalty terms. Together, these results address a central question: \\
\textit{To what extent can pure and hybrid quantum annealing already deliver viable solutions to stochastic optimization problems of practical importance in power system operations?}

Our findings show that while current annealing hardware cannot yet embed realistic stochastic UCP instances, the hybrid solver is capable of producing feasible solutions close to classical benchmarks under practical runtime limits. Moreover, Simulated Annealing (SA) experiments on the QUBO reformulation highlight the potential of annealing-based heuristics when supported by adaptive penalty tuning. Together, these results establish a methodological foundation for future quantum-inspired approaches to power system optimization under uncertainty.

\begin{figure}
    \centering
    \includegraphics[width=0.7\linewidth]{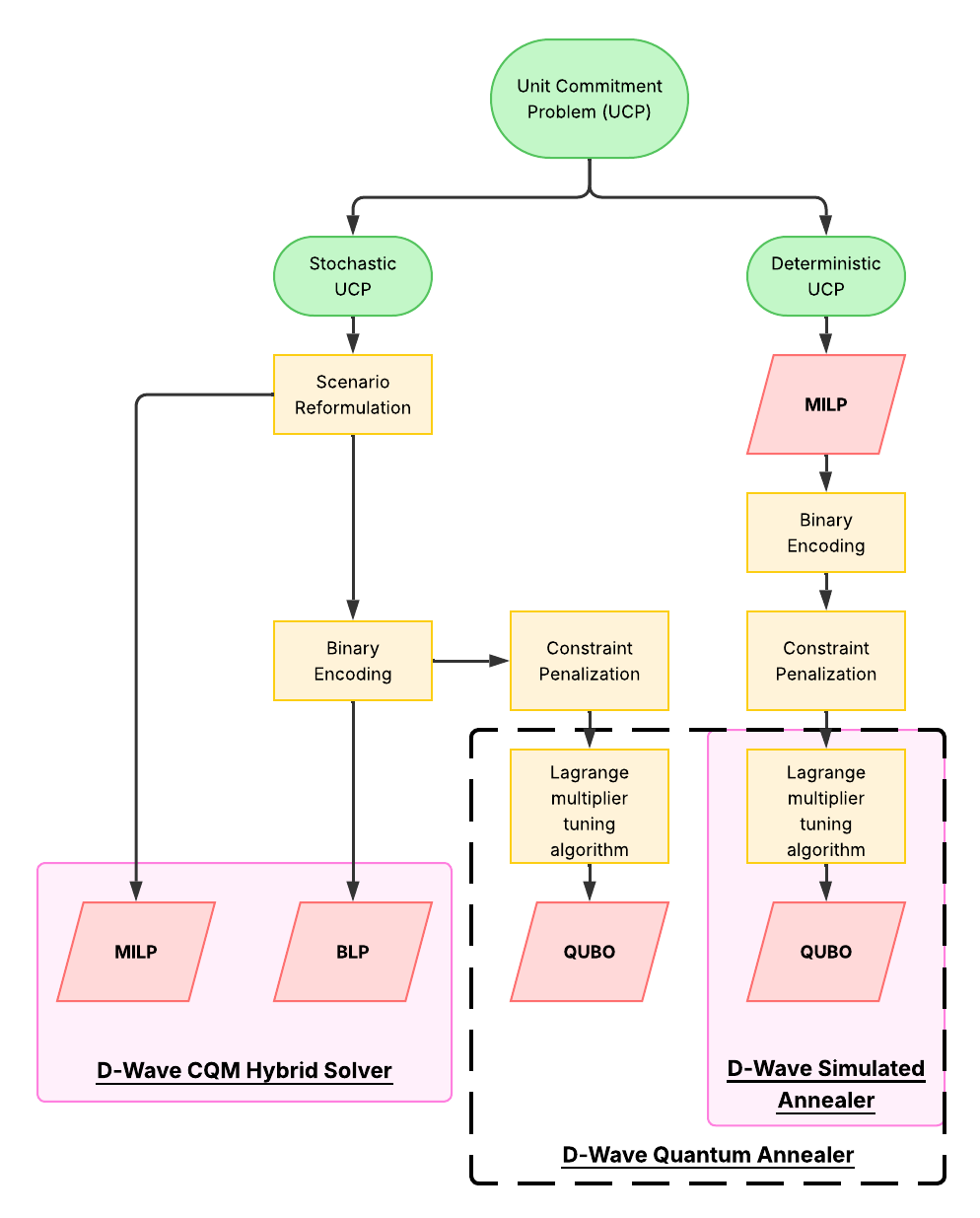}
    \caption{Flow diagram of reformulations of deterministic and stochastic UCPs into different instances. Both problems in their MILP formulation were solved using Gurobi for benchmarking purposes. Black dashed box: QUBOs unsolved on D-Wave due to size limitations (see Section \ref{sec:results_det_formualtion}).}
    \label{fig:flow_diagram}
\end{figure}

The structure of this work is as follows. In Section \ref{sec:problem_formulation} we acquaint chance constrained programming and introduce chance constraints in a UCP. Section \ref{sec:results} shows the results of every instance investigated given by the different solvers used (see Figure \ref{fig:flow_diagram} for a flow diagram of every instance and solver used). A discussion is presented in Section \ref{sec:discussion}, and the methodology of all reformulations carried out throughout the work is presented in Section \ref{sec:methods}.

\section{Chance-Constrained Unit Commitment Problem}\label{sec:problem_formulation}

\subsection{Background: Chance Constrained Programming}

\emph{Chance-constrained programming}, introduced by Charnes and Cooper \cite{Charnes-Cooper1959}, is a cornerstone of stochastic optimization. Unlike deterministic programs, whose constraints must hold exactly, chance-constrained programs require that restrictions involving uncertain quantities hold with a specified probability. For detailed treatments, see \cite{Prekopa1995, Prekopa2003, Shapiro2009}. Applications range from aeronautics \cite{Caillau2018} and gas transport \cite{Henrion2025, Schuster2022} to population dynamics \cite{PerezAros2022}. Within the energy domain, related studies include economic \cite{Loaiciga1988, PrekopaSantai1978, vanAckooij2014, Berthold2022} and power dispatch problems \cite{Ordieres2021, Hong2022, Ouanes2024}. Despite this extensive body of work, \emph{joint chance-constrained programs} have yet to be explored on quantum platforms.

A program is said to be chance-constrained when it takes the form
\begin{equation}
\begin{aligned}
\min_{\mathbf{x}} \quad & f(\mathbf{x}) \
\text{s.t.} \quad & \mathbb{P}\left(g(\mathbf{x},\boldsymbol{\xi}) \le 0\right) \ge p,
\end{aligned}
\end{equation}
where $\mathbf{x}\in\mathbb{R}^n$ is the decision vector, $\boldsymbol{\xi}\in\mathbb{R}^m$ a random vector, and $p\in(0,1)$ the required reliability level. Here $f: \mathbb{R}^n \to \mathbb{R}$ denotes the objective, and $g: \mathbb{R}^n \to \mathbb{R}^k$ encodes the constraints under uncertainty. When $k=1$, the model involves an \emph{Individual Chance Constraint} (ICC); for $k\ge2$, a \emph{Joint Chance Constraint} (JCC), requiring all conditions to hold simultaneously.
Formally, let $\boldsymbol{\xi}$ be an $m$-dimensional random vector admitting a density $f_{\xi}$ with respect to Lebesgue measure $\lambda_m$ on $\mathbb{R}^m$. For a decision parameter $\mathbf{x}$, the probability of simultaneous satisfaction of the inequalities can be written as
\begin{equation}
    \varphi(\mathbf{x}) = \int_{\{z \in \mathbb{R}^m : g(x,z)\leq 0\}} f_\xi(z)\, d\lambda_m(z) 
    = \mathbb{P}\left(g(\mathbf{x},\boldsymbol{\xi}) \leq 0\right).
\end{equation}

Unlike ICCs, which require each constraint to hold with a given probability independently, JCCs demand all inequalities to be satisfied simultaneously with high probability. This introduces a multidimensional probability measure over the feasible set in $\mathbb{R}^m$. The joint distribution of $\boldsymbol{\xi} = (\xi_1,\ldots,\xi_m)$ is assumed to be known; otherwise, the probabilistic constraint in (2) would not be well defined.

Numerically, evaluating or approximating $\varphi(\mathbf{x})$ remains challenging. Two main classes of methods have emerged. \emph{Discretization-based approaches}, employed here,  approximate the distribution of $\boldsymbol{\xi}$ through finite sampling: via scenarios, Monte Carlo, or quasi–Monte Carlo schemes \cite{Kleywegt2002, CalafioreCampi2006, CampiGaratti2008, Shapiro2009}. The resulting deterministic reformulations are tractable but sensitive to the number and selection of samples \cite{CampiGaratti2018, Ahmed2008}. In contrast, \emph{continuous approximation methods} exploit analytic or structural features of the distribution and constraints, preserving the continuous character of the problem but often requiring stronger assumptions on $f_\xi$ and $g$ \cite{Prekopa1995, Shapiro2009, Henrion2025, Grandon2022, Ahmed2008}.

\subsection{Problem Formulation: Stochastic Chance-Constrained UCP}

We study a stochastic UCP formulated as a MILP with JCCs, ensuring system demand is met with probability at least \(p \in (0,1)\). The deterministic base model is adopted from \cite[Ch.~7]{Conejo}.

Let \(u_{g,t} \in \{0,1\}\) denote the on/off status of generator \(g \in \{1,\dots,G\}\) at time \(t \in \{1,\dots,T\}\) (note $G = T = 3$ for our case). Let \(z_{g,t}^{\text{on}}, z_{g,t}^{\text{off}} \in \{0,1\}\) denote start-up and shut-down indicators, respectively, and let \(p_{g,t} \in \mathbb{R}_{+}\) denote the generation output. The stochastic UCP reads: 

\begin{alignat}{2}
\label{eq:objective}
\min \quad &
\sum_{g=1}^{G} \sum_{t=1}^{T} \Big(
    z_{g,t}^{\text{on}} C^{\text{StartUp}}_{g} 
    + z_{g,t}^{\text{off}} C^{\text{ShutDown}}_{g} 
    + C_g u_{g,t} 
    + b_g p_{g,t}
\Big) \\
\text{s.t.} \quad 
\label{eq:logic}
& u_{g,t} - u_{g,t-1} = z_{g,t}^{\text{on}} - z_{g,t}^{\text{off}}
&\quad  &\forall g,t \\
\label{eq:exclusive}
& z_{g,t}^{\text{on}} + z_{g,t}^{\text{off}} \leq 1
&\quad & \forall g,t \\
\label{eq:capacity}
& P_{g}^{\min} u_{g,t} \leq p_{g,t} \leq P_{g}^{\max} u_{g,t}
&\quad & \forall g,t \\
\label{eq:ramp}
& -R^{\text{down}}_{g} \leq p_{g,t} - p_{g,t-1} \leq R^{\text{up}}_{g}
&\quad & \forall g,t \\
\label{eq:prob}
& \mathbb{P}\!\left(
    D_t \leq \sum_{g=1}^{G} p_{g,t}, \ \forall t
\right) \geq p \\
& u_{g,t},\, z_{g,t}^{\text{on}},\, z_{g,t}^{\text{off}} \in \{0,1\}, \quad
  p_{g,t} \in \mathbb{R}_{+}. \nonumber
\end{alignat}

Parameters about the generators, the demand, and the initial commitment status are specified in Tables \ref{tab:miqpparameters_values_stochastic}, \ref{tab:demands_stochastic} and \ref{tab:priors_stochastic}, respectively, from Appendix \ref{app:parameters_stoch_UCP}. The stochastic demand $\mathbf{D} = (D_1,\ldots,D_T)^{\top}$ satisfies
\[
\mathbf{D} \sim \mathcal{N}(\mu,\Sigma),
\]
with mean vector $\mu \in \mathbb{R}^T$ and covariance matrix $\Sigma \in \mathbb{R}^{T \times T}$. The dependence structure is specified through the correlation matrix
\[
R = (\rho_{st})_{s,t=1},
\]
with $\Sigma = \mathrm{diag}(\boldsymbol{\sigma})\, R \,\mathrm{diag}(\boldsymbol{\sigma})$, where $\sigma_t = \sqrt{\Sigma_{tt}}$ are the marginal standard deviations. Table~\ref{tab:correlations} reports exemplary correlation regimes for successive time periods.

The constraints \eqref{eq:logic}-\eqref{eq:ramp} from the optimization problem \eqref{eq:objective} enforce consistency of commitment and switching decisions, generation capacity limits, and ramping restrictions. The JCC \eqref{eq:prob} imposes that scheduled generation exceeds demand simultaneously for all time periods with probability at least $p$. Such probabilistic constraints are, in general, intractable in MILP optimization and necessitate approximation as shown in section \ref{sec:stochastic_UCP}.

\begin{table}[h]
\centering
\begin{tabular}{c|c|c|c|}
\cline{2-4}
 & $\rho_{12}$ & $\rho_{13}$ & $\rho_{23}$ \\ \hline
\multicolumn{1}{|c|}{No correlations} & 0 & 0 & 0 \\ \hline
\multicolumn{1}{|c|}{Moderate correlations} & 0.3 & 0.4 & 0.5 \\ \hline
\multicolumn{1}{|c|}{Strong correlations} & 0.6 & 0.7 & 0.8 \\ \hline
\end{tabular}
\caption{Correlations regimes for the demand vector $\mathbf{D}$.}
\label{tab:correlations}
\end{table}

\section{Results}\label{sec:results}

This section reports the results obtained across all test instances using three solvers:

\begin{itemize}
\item[(i)] \textit{Gurobi Optimizer}: a state-of-the-art solver suited for mixed-integer programming, used here as the classical benchmark. The computations were carried out on a dedicated computer cluster equipped with two 2.4 GHz Intel Xeon Gold 5115 CPUs (10 cores each) and 96 GB of RAM;
\item[(ii)] \textit{D-Wave hybrid CQM solver}:  a quantum–classical approach combining heuristic search with quantum annealing on the Advantage2 QPU;
\item[(iii)] \textit{D-Wave simulated annealer}: a classical heuristic for QUBO formulations.
\end{itemize}

Each subsection specifies the solver employed: all results are compared against Gurobi; the stochastic UCP is addressed with the hybrid solver; and the deterministic UCP with the simulated annealer.

\subsection{Performance of Classical and Hybrid Solvers for the Stochastic UCP}

The scenario reformulation of the stochastic MILP with joint chance constraints (see details in section \ref{sec:stochastic_UCP}), with constraints from equations \eqref{eq:objective}-\eqref{eq:ramp},\eqref{eq:scenario_balance},\eqref{eq:scenario_reliability},
is solved under three correlation regimes of the Gaussian demand model (see Table~\ref{tab:correlations}). All results in this section are obtained with Gurobi and D-Wave's hybrid quantum-classical solver.

\begin{figure}
    \centering
    \includegraphics[width=\linewidth]{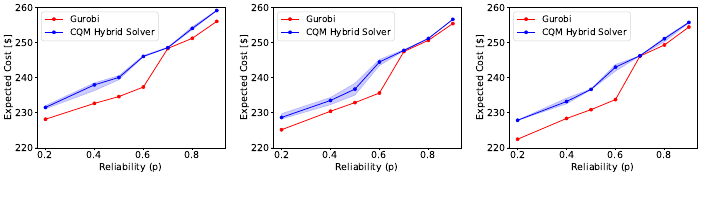}
    \caption{Expected cost of the MILP reformulated stochastic UCP as a function of the reliability level $p$ of the joint chance constraints, 
    comparing Gurobi and D-Wave’s hybrid CQM solver. Subfigures correspond to: (a) uncorrelated, (b) moderately correlated, 
    and (c) strongly correlated Gaussian demand distributions. For each instance, 1000 scenarios are considered, and each run has a runtime limit of 20 seconds. For the hybrid solver, results are based on 5 runs, each of them producing an average of about 100 solutions; the reported values correspond to the average of the 5 lowest costs of those runs, with the shaded region indicating the range from best to worst among them. Each D-Wave run included 
    32 ms of QPU access time.}
    \label{fig:expected_costs}
\end{figure}

Figure~\ref{fig:expected_costs} displays the expected generation cost as a function of the reliability level~$p$. 
As anticipated, costs increase monotonically with~$p$, since higher reliability requires more conservative scheduling. 
This monotonic trend is consistent across all correlation regimes and reflects the inherent trade-off between reliability 
and cost in chance-constrained optimization. A further systematic effect is observed with respect to correlations: 
the higher the positive correlation among demands, the lower the resulting cost. Positive dependence reduces the dispersion 
of the aggregate demand and thereby decreases the conservatism required in scheduling decisions.

Across all instances, Gurobi attains lower expected costs than the hybrid CQM solver. 
The hybrid solutions, however, remain close and in several cases nearly coincide with those of Gurobi. 
The shaded regions around the CQM curves indicate variability between repeated runs: each value shown 
is the mean of the five lowest-cost solutions, while the band spans the best to worst among them. 
This variability reflects the heuristic sampling nature of the hybrid method, in contrast to the deterministic 
behavior of Gurobi. 

The cost gap between solvers is largest at low reliability levels, where the feasible set is wide. From about $p=0.7$ upwards, the feasible set becomes sufficiently restricted for the hybrid solver to identify near-optimal schedules, and the expected costs of both solvers are almost identical. The precise threshold at which this alignment occurs depends on the distributional assumptions and the scenario sampling, and in these experiments is observed around $p=0.7$.

These results demonstrate that scenario reformulations of the UCP with JCCs can be solved 
on the hybrid quantum--classical platform with solution quality close to that of a state-of-the-art MILP solver. 
The discrepancy narrows as reliability requirements increase, indicating particular promise of hybrid methods 
in planning settings where high reliability is critical.

\subsection*{Scenario Sample Scaling}

Here we examine how the size and performance of the scenario reformulated MILP and constrained Binary Linear Program (BLP) formulations (see Section \ref{sec:stochastic_UCP}) vary with the number of sampled scenarios. The solvers used are Gurobi and D-Wave’s hybrid quantum–classical solver. Correlations between demands at different time periods are set to be moderate (see Table \ref{tab:correlations}). Sample sizes range from 800 to 20,000, fixing the stochastic dimension while refining the discrete approximation.

In the scenario reformulation presented in in Section \ref{sec:stochastic_UCP} it is shown that each sampled scenario contributes additional constraints of type \eqref{eq:scenario_balance}, so that the number of constraints grows proportionally to $N T$, where $T$ denotes the number of time periods and $N$, the number of scenarios considered. Also, a new set of binary variables $y_i$ is introduced, which increases linearly with $N$, while the original unit commitment variables contribute a term of order $G T$, with $G$ denoting the number of generators. In the BLP, the number of variables increases further due to binary encoding: each continuous generation variable $p_{g,t}$ is replaced by $n = \left\lceil \log_{2} \left( \max_{g} \left\{ P_g^{\text{max}} - P_g^{\text{min}} \right\} + 1 \right) \right\rceil$ binary variables, which is the minimum number of variables needed to represent the whole range of integer numbers of the continuous variables (see Appendix \ref{app:slack_number}). Hence, the total number of decision variables grows proportionally to $G T \log_2(P_g^{\max}-P_g^{\min})$.

As the number of scenarios $N$ increases, the feasible set of the sampled problem converges almost surely to the true feasible set of the chance-constrained problem \cite[Ch.~5--6]{Shapiro2009}.  Accordingly, the optimal values of the sampled problems converge to the true optimal value. For finite sample sizes, however, the feasible sets are random and no monotonicity can be guaranteed. In practice, small sample sizes may yield pessimistic (overestimated) costs when minimizing, which tend to be corrected as the sample size grows.

\begin{figure}
    \centering
\includegraphics[width=0.55\linewidth]{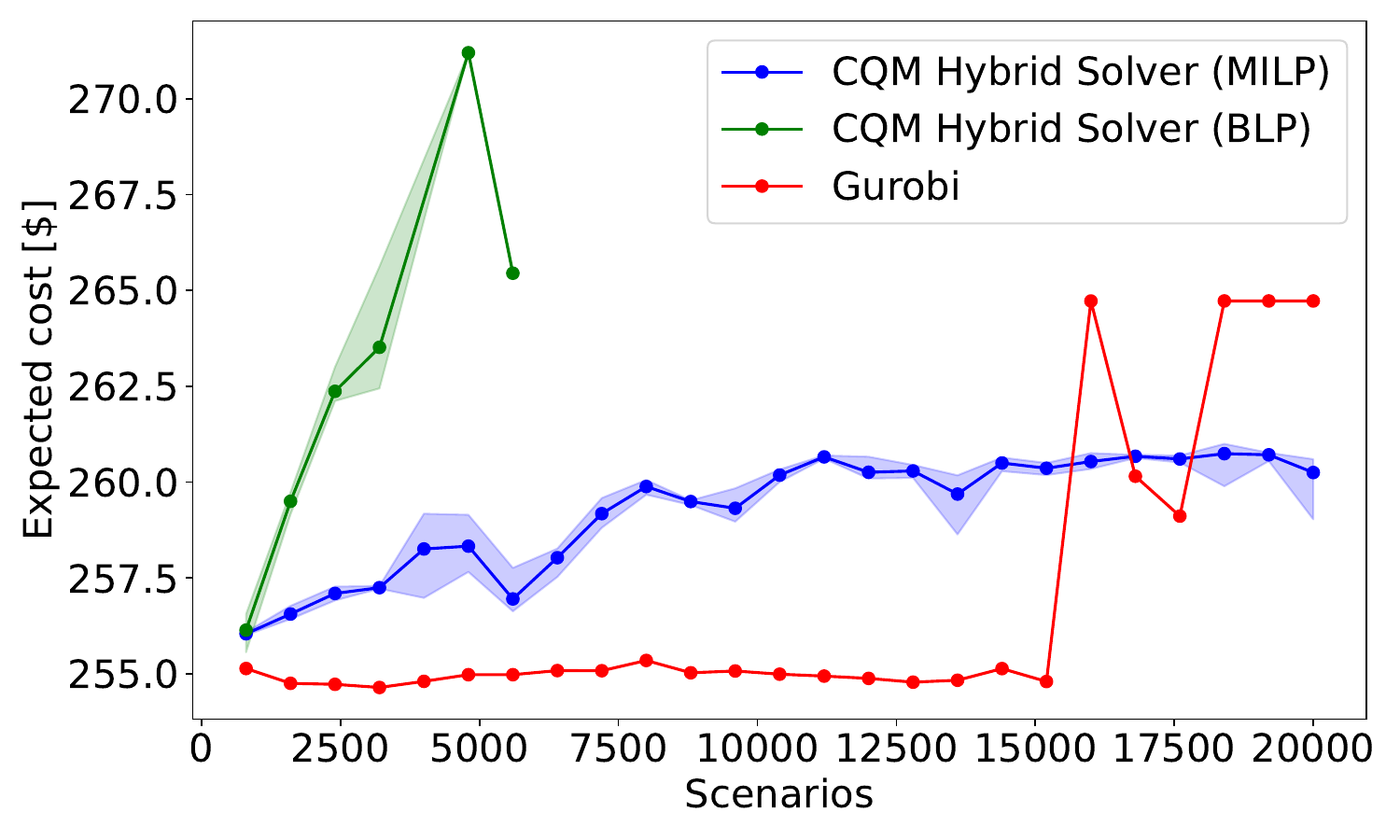}
   \caption{Expected cost of the stochastic UCP with moderate correlations as a function of the number of scenarios. 
    Results compare Gurobi with D-Wave’s hybrid solver (MILP and BLP formulations). The shaded areas correspond to error bars indicating variability 
    across runs; missing points for the BLP reflect the absence of feasible solutions. Each run has a runtime limit of 35 seconds. For the hybrid solver, results are based on 4 runs, each of them producing an average of about 100 solutions. The QPU access time of the hybrid solver ranges from 0ms to 70ms (see Section \ref{sec:discussion} and Figure \ref{fig:QPU_time} for more details).}
    \label{fig:scalability}
\end{figure}

To enable a fair comparison between classical and hybrid methods, both solvers were run under the same time budget of 35 seconds per instance. Figure~\ref{fig:scalability} shows the expected costs for increasing scenario counts. For moderate sample sizes, we anticipated that Gurobi would decrease expected costs with increasing sample sizes, since poor discretization yields pessimistic results; however, costs remain mostly constant as the number of scenarios grows and the approximation becomes more accurate. Beyond 15,000 scenarios, Gurobi produces higher costs within the fixed time limit, whereas the hybrid solver continues to return feasible solutions of comparatively lower cost. These results indicate that the hybrid method may be more effective in large-scale settings under tight runtime constraints. 

The BLP formulation (green line), by contrast, ceases to yield feasible solutions once the number of scenarios exceeds approximately 6,000. At first, one might suspect that this behavior is due to the linearly increasing number of binary variables introduced by the encoding of continuous variables. However, this explanation is not consistent with the MILP formulation (blue line), which remains capable of producing feasible solutions even as the number of scenarios, and consequently the number of variables, grows beyond that of smaller BLP instances.

A more likely source of difficulty lies in the binary encoding itself. In particular, if the binary part of the problem is processed by a classical metaheuristic such as simulated annealing, the neighborhood structure may be poorly matched to the encoding: single-bit flips can correspond to large changes in the underlying continuous value, while small changes may require several simultaneous flips. Such a mismatch can prevent effective exploration of the solution space and lead to infeasibility in larger instances. This observation suggests that, for the binary subproblem, the hybrid solver may rely, at least partly, on classical rather than quantum resources. The precise reasons for the failure of the BLP formulation, however, remain open.

\begin{figure}
    \centering
    \includegraphics[width=0.5\linewidth]{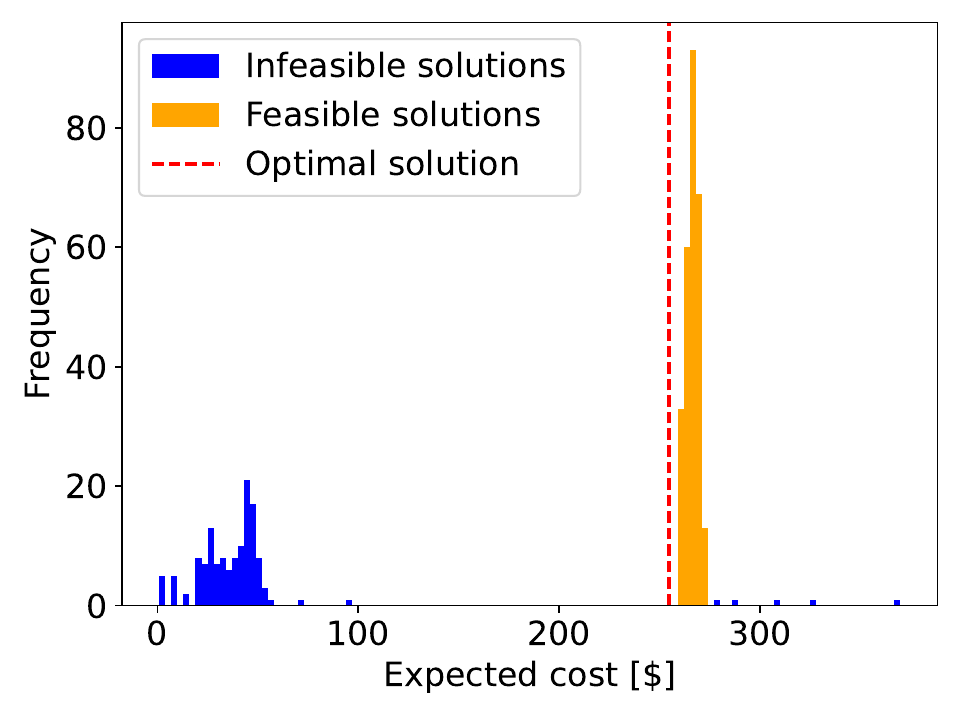}
     \caption{Feasibility histogram of the stochastic UCP with 15,200 scenarios in its MILP formulation solved on D-Wave’s hybrid solver. The solver gives 404 solutions, 268 are feasible (orange) and 136 are infeasible (blue). Feasible solutions concentrate near the optimal cost of 254.8 \$, while a large number of infeasible solutions are also returned. The best feasible solution found has expected cost 260.2 \$.}
    \label{fig:histogram}
\end{figure}

Feasibility of returned solutions by the hybrid solver in the MILP formulation represents a further important aspect of solver performance. Figure~\ref{fig:histogram} displays the distribution of solutions obtained with the hybrid solver 
for an instance with 15,200 scenarios in the MILP formulation. Feasible solutions (orange) occur with non-negligible frequency, with objective values concentrated in a narrow band close to the best-known stochastic optimum (254.8 \$). 
The best feasible solution attains a cost of 260.2 \$, corresponding to a gap of roughly 2\%. At the same time, a substantial fraction of solutions are infeasible, with costs widely dispersed at lower values. This reflects a potential use of penalty-based methods by the hybrid solver, and the difficulties of such methods to navigate the feasible set: feasibility is not guaranteed but depends on appropriate calibration of penalty factors.

Overall, these results underline the dual challenge of stochastic quantum optimization: not only to approximate 
expected costs, but also to ensure feasibility under probabilistic constraints. The hybrid solver is capable of 
producing near-optimal feasible solutions, yet infeasibility remains a frequent outcome for large scenario sets.

\subsection{QUBO Reformulation: Complexity and Embedding Limits}\label{sec:results_det_formualtion}

To analyze to what extent a pure quantum annealer can be used to solve a stochastic optimization problem with joint chance constraints, we have converted the original MILP, see Equations \eqref{eq:objective}–\eqref{eq:ramp} and Equations \eqref{eq:scenario_balance}, \eqref{eq:scenario_reliability} from Section \ref{sec:stochastic_UCP}, into a QUBO problem. To do that, we binary encode continuous variables and introduce penalty terms in the cost function to enforce the constraints. The procedure follows exactly the same steps as those detailed in section \ref{sec:deterministic_UCP}, where it is illustrated for a deterministic UCP; the principles carry over unchanged to the stochastic case considered here.

All decision variables are binary encoded, and both operational and chance constraints are represented by quadratic penalty terms. The total number of binary variables increases with problem size according to the following components:
\begin{itemize}
  \item[a)] unit commitment and switching variables, $z^{\text{on}}_{g,t}$, $z^{\text{off}}_{g,t}$ and $u_{g,t}$, of order $3GT$, 
  \item[b)] binary encodings of continuous generation levels, $p_{g,t}$, of order $GTn$, where $n$ denotes the number of bits required per continuous variable, 
  \item[c)] scenario indicators, $y_i$, of order $N$, where $N$ is the number of sampled scenarios, 
  \item[d)] slack variables introduced for inequality constraints. Each sampled scenario contributes four such inequalities (demand, reliability, ramping limits and maximum capacity). In the QUBO reformulation, each inequality requires a slack variable binary encoded with $N_s$ binary variables. The $N_s$ values for each inequality can be computed following the procedures presented in Appendices \ref{app:slack_number}, \ref{app:slack_det_demand}, \ref{app:slack_ramp}, \ref{app:slack_capacity}. For $N_s^{\text{reliability}}$ see Section \ref{sec:stochastic_UCP}.
\end{itemize}

Hence the total number of binary variables is given by
\begin{equation}
3GT + GTn + N + N_s^{\text{ramp}} + N_s^{\text{reliability}} + N_s^{\text{demand}} + N_s^{\text{capacity}}, 
\label{number}
\end{equation}
and thus scales linearly with the number of generators $G$, time periods $T$, bit depth $n$, number of scenarios $N$, and slack-variable encodings $N_s$.

Even for modest instances, the QUBO formulation of the stochastic UCP leads to substantial binary dimensions. For example, with $G = T = 3$ and $N = 1000$, the problem involves approximately $5\times10^4$ binary variables. This rapid growth reflects the combined contributions of commitment and switching variables, binary encodings of continuous generation levels, scenario indicators, and the slack variables required to represent inequality constraints. The presence of chance constraints further increases model density by inducing quadratic couplings that link variables across multiple scenarios.

\begin{figure}
    \centering
\includegraphics[width=0.7\linewidth]{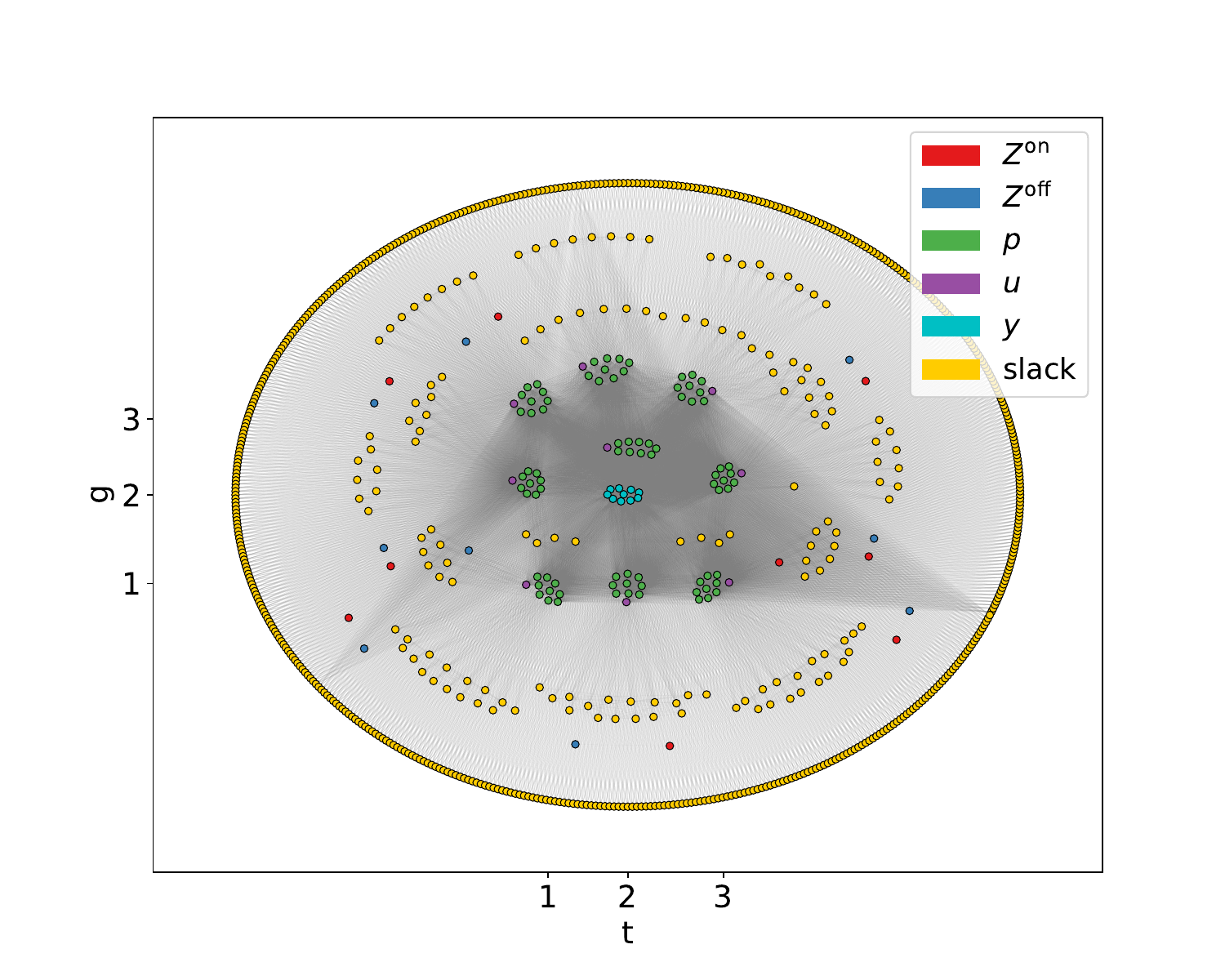}
    \caption{Induced QUBO graph for the stochastic UCP considering 10 scenarios. Each node corresponds to a binary decision variable and edges denote quadratic couplings arising from the cost function. The slack variables forming a circle around the graph correspond to those used for the demand inequality constraints.}
    \label{fig:graph_stoch_UCP}
\end{figure}

The resulting QUBO graph becomes dense even for small $N$. Figure~\ref{fig:graph_stoch_UCP} illustrates the case $N=10$, where the slack variables associated with the demand constraints \eqref{eq:scenario_balance} are positioned around the periphery. This layout emphasizes their structural role: through quadratic penalty terms enforcing demand satisfaction, each slack variable couples to numerous binary-encoded generation variables and scenario indicators.These interactions form a highly connected graph in which peripheral slack nodes are densely linked to a central core of generation and commitment variables.

Such dense connectivity presents significant challenges for quantum annealing hardware. The combination of high variable counts and largely connected graphs is incompatible with the sparse, fixed topologies of current quantum devices. The D-Wave Advantage2 system, for instance, provides roughly $4{,}400$ qubits with a maximum connectivity of 20 \cite{Dwave_whitepaper}. Embedding dense QUBOs into this architecture requires constructing chains of physical qubits to represent single logical variables, increasing susceptibility to errors and compounding the effects of hardware noise \cite{Dwave2020}. As a result, even relatively small stochastic QUBOs with joint chance constraints exceed the embedding and noise-tolerance capabilities of existing annealers, limiting feasible experiments to simplified deterministic or small-scale instances.

\subsection*{Deterministic UCP}

\begin{figure}
    \centering
\includegraphics[width=0.7\linewidth]{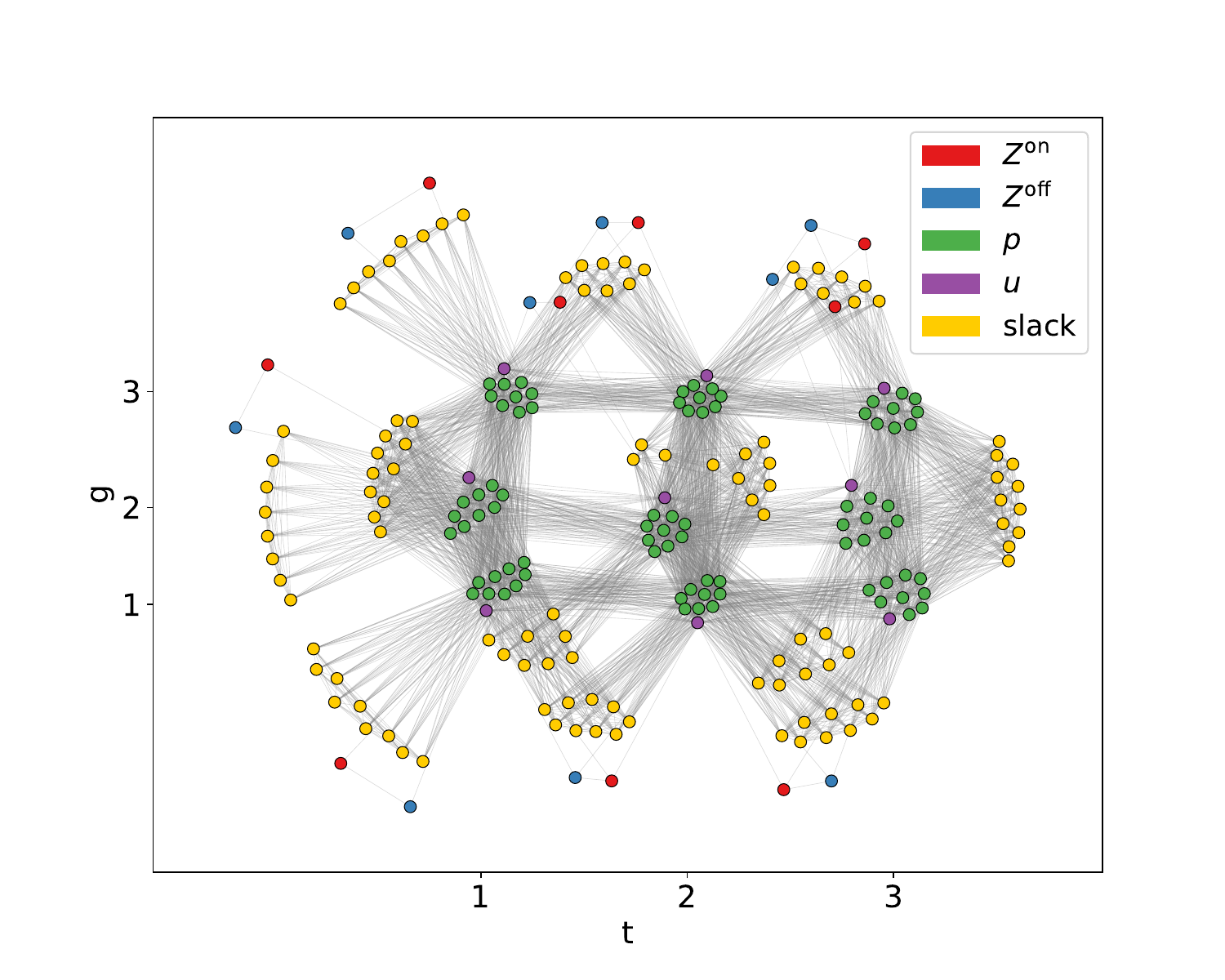}
    \caption{Induced QUBO graph for the deterministic UCP. Each node corresponds to a binary decision variable and edges denote couplings arising from penalized constraints and quadratic terms in the objective function.}
    \label{fig:qubo_deterministic}
\end{figure}

Given these hardware constraints, we next restrict our attention to the deterministic UCP, where demand values are fixed rather than random (see Table \ref{tab:demands_deterministic} in Appendix \ref{app:parameters_stoch_UCP}). This reduction substantially decreases model complexity: for $N=10$, the stochastic QUBO comprises 809 binary variables and 26{,}781 quadratic couplings, whereas the deterministic formulation contains only 291 variables and 5{,}651 couplings (see Figure~\ref{fig:qubo_deterministic}). The dense connectivity illustrates the embedding complexity even in the absence of uncertainty.

At first sight, such dimensions may seem compatible with the specifications of D-Wave’s Advantage2 system \cite{Dwave_whitepaper}. However, feasibility of embedding depends not only on the number of variables but also on the number of quadratic terms and the structure of the interaction graph. The Zephyr topology of Advantage2 \cite{Dwave_topology} is designed for graphs with relatively homogeneous degree distributions. By contrast, the QUBO form of the deterministic UCP produces interaction graphs with highly uneven connectivity, where some variables are largely clustered. Embedding such graphs still requires long chains of physical qubits, which are error-prone and increase sensitivity to hardware noise.

Thus, even in the deterministic setting, the QUBO formulation of the UCP cannot be embedded 
effectively on the available quantum annealing hardware.
In view of this limitation, we employ simulated annealing as a proxy for quantum annealing. 
Simulated annealing is a classical metaheuristic designed for the minimization of functions 
over binary variables \cite{Pincus1970}. Starting from an initial configuration (randomly chosen 
in our experiments), the method iteratively proposes single-bit flips. Each proposed move is 
accepted or rejected with a probability that depends on the induced change in the objective 
function and on a decreasing temperature parameter. This mechanism allows occasional acceptance 
of worsening moves, enabling the algorithm to escape local minima and to explore the energy 
landscape more broadly. In theory, convergence to a global optimum is guaranteed under an 
infinitely slow cooling schedule, but in practice the algorithm may become trapped in local 
optima due to cooling time limits \cite{Guilmeau2021}. 

As a purely classical approach, simulated annealing imposes no restrictions related to hardware 
embedding. At the same time, the QUBO formulation developed here remains fully compatible with 
future generations of quantum annealers, whose improved connectivity may allow the embedding of 
larger problem graphs. For reproducibility, the source code used is publicly 
available \cite{Ribes2025}.

Using simulated annealing, we obtain feasible solutions within $8$--$10\%$ of the optimum, and 
observe systematic improvements in feasibility ratios when penalty factors are tuned. These findings suggest that while 
current hardware cannot yet accommodate realistic UCP QUBOs, annealing-based heuristics already 
provide a meaningful first step toward quantum-inspired methods for stochastic power system optimization.  

For the QUBO formulation of the deterministic UCP, the penalty parameters associated with constraint violations were tuned using the algorithm described in section \ref{methods:lagrange_tuning}. Our novel algorithmic procedure is an adaptive scheme for tuning penalty factors in QUBO formulations, where the weights are iteratively updated based on observed feasibility ratios of sampled solutions, thereby steering the search toward constraint satisfaction without resorting to excessively large penalties.

\begin{figure}
    \centering
    \includegraphics[width=\linewidth]{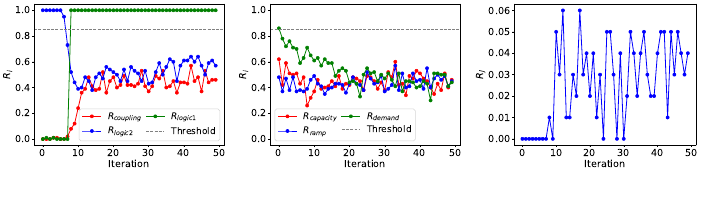}
    \caption{Evolution of feasibility ratios $R_i$ and overall $R_J$ during penalty factors updates. 
    Results are shown for (a) binary (logic, coupling), (b) continuous (capacity, demand, ramping), 
    and (c) joint feasibility. Parameters of the sigmoid update rule: $k=14$, $R_0=0.3$, $A=0.5$ (see Section \ref{methods:lagrange_tuning} for details on these parameters).}
    \label{fig:feasibility_evolution}
\end{figure}

\begin{table}[H]
\centering
\begin{tabular}{c|c|c|c|c|c|c|}
\cline{2-7}
 & $\lambda_{\text{logic1}}$ & $\lambda_{\text{logic2}}$ & $\lambda_{\text{demand}}$ & $\lambda_{\text{coupling}}$ & $\lambda_{\text{capacity}}$ & $\lambda_{\text{ramp}}$ \\ \hline
\multicolumn{1}{|c|}{Value} & 24.62 & 3.63 & 7.21 & 1081.48 & 31.61 & 37.32 \\ \hline
\end{tabular}
\caption{Final penalty factors after 50 iterations of our proposed update scheme.}
\label{tab:Lagrange_multipliers}
\end{table}

Let $R_i$ denote the feasibility ratio of solutions satisfying constraint $i$, 
and $R_J$ the joint feasibility ratio (all constraints satisfied simultaneously).
Figure~\ref{fig:feasibility_evolution} reports the evolution of $R_i$ and $R_J$ over 50 iterations. 
While overall feasibility remains limited, with roughly $3\%$ of solutions jointly feasible, 
the update scheme stabilizes the ratios and prevents divergence of the penalty factors. 
The final set of penalty values is listed in Table~\ref{tab:Lagrange_multipliers}. 

\begin{figure}
    \centering
    \includegraphics[width=0.5\linewidth]{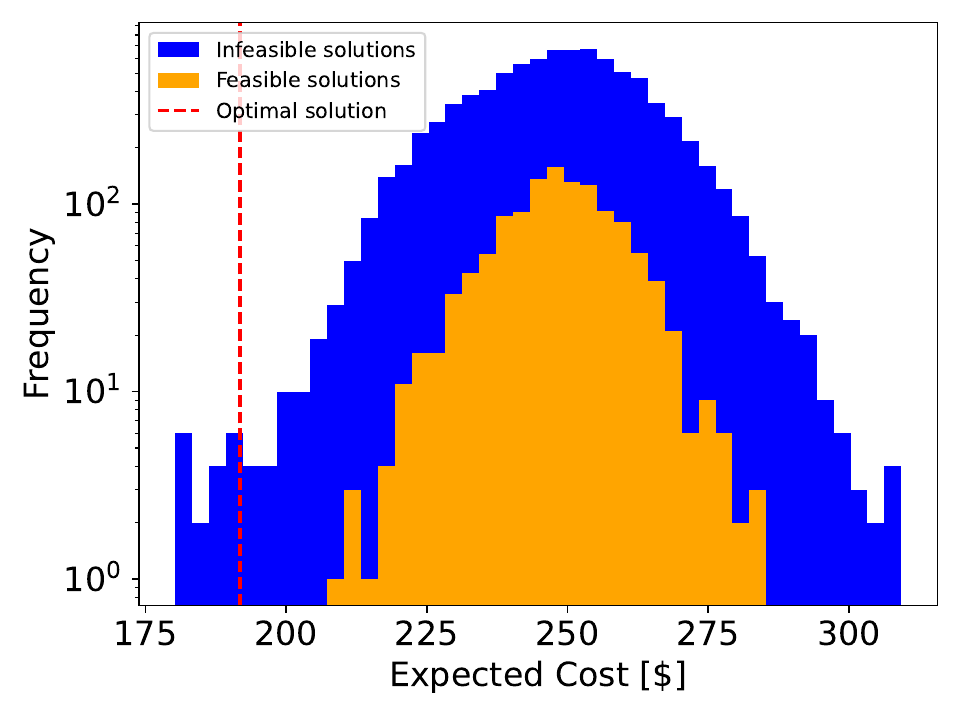}
    \caption{Feasibility histogram of solutions of the deterministic UCP (formulated as a QUBO program) obtained with D-Wave's simulated annealer. The results consist of a single run with $10^4$ reads. The best feasible solution is 208.75$ \$$, and the optimal solution is $191.8 \$$.}
    \label{fig:feasibility_BQM}
\end{figure}

Once the penalty factors were fixed, the QUBO was solved by simulated annealing. 
From a single run with $10^4$ reads, both feasible and infeasible solutions were obtained. 
Figure~\ref{fig:feasibility_BQM} shows the resulting distribution of objective values. 
Most solutions concentrate around costs of approximately 250 \$, with feasible solutions 
forming a narrower subset. The best feasible solution achieved a cost of 208.75 \$, 
corresponding to a deviation of 8.8\% from the deterministic optimum, while the average 
gap across feasible solutions was closer to 30\%.

This performance gap can be attributed to the search dynamics of simulated annealing, which relies on 
single-bit flips and is therefore prone to becoming trapped in local minima on problems relying on a binary encoding. A genuine quantum annealer, 
operating under a fundamentally different search mechanism, would not be subject to these same 
limitations. These results therefore underline both the usefulness of the tuning scheme for improving 
feasibility and the potential of future quantum hardware with larger qubit counts and richer connectivity 
to narrow the gap with classical optimization solvers.

\section{Discussion}\label{sec:discussion}

This work provides, to our knowledge, the first systematic investigation of quantum annealing for 
chance-constrained unit commitment models. Figure~\ref{fig:flow_diagram} summarizes our methodological 
workflow: from MILP to binary encodings and QUBO formulations executed on D-Wave’s hybrid and simulated annealers. This unified 
architecture enabled assessment of the tractability and performance of classical, hybrid, and quantum 
approaches to stochastic power system optimization.

The scenario approximation of the chance-constrained UCP served as a tractable testbed for comparing 
solver behavior under uncertainty. The hybrid quantum–classical solver was capable of handling very large 
scenario sets—up to 15,000 in our experiments, whereas Gurobi’s branch-and-bound method became increasingly 
burdened by combinatorial growth. Under strict runtime limits, the hybrid solver achieved solution quality 
comparable to Gurobi, suggesting that hybrid schemes may hold promise for time-constrained, large-scale 
stochastic settings. Whether this advantage persists under relaxed time limits or full optimality conditions 
remains an open question.

\begin{figure}
    \centering
    \includegraphics[width=0.5\linewidth]{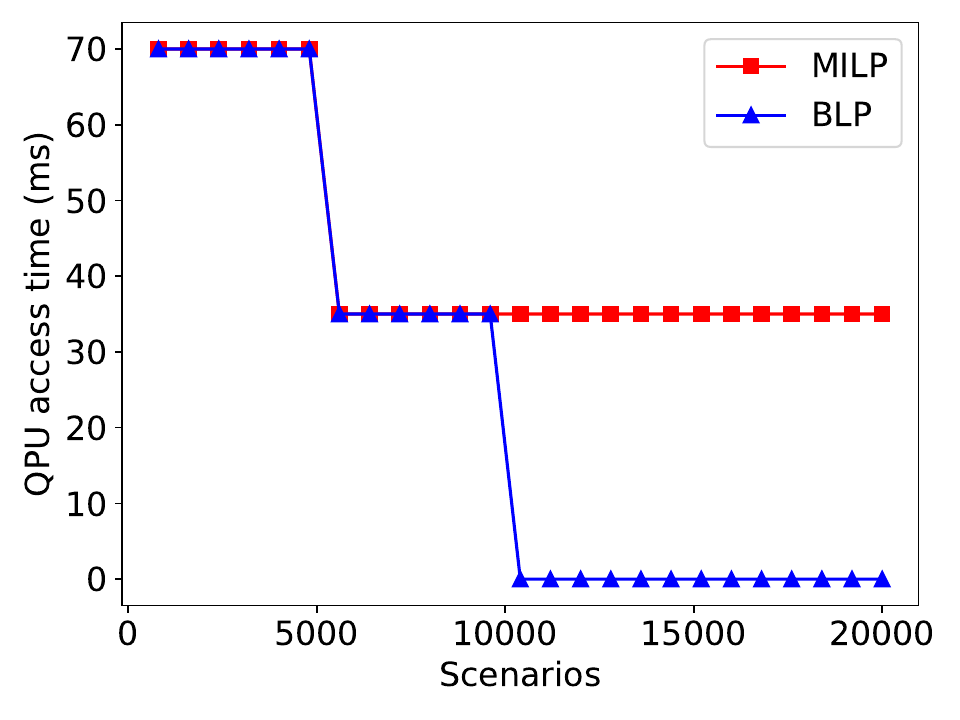}
    \caption{QPU access time of the hybrid solver comparison between the MILP and the BLP instances, as a function (i.e. complexity) of the number of scenarios drawn from the multivariate normal distribution sampleset.}
    \label{fig:QPU_time}
\end{figure}

A central limitation of the D-Wave hybrid solver is its opacity: the interaction between quantum and classical 
components, including convergence and resource scheduling, is proprietary and undisclosed. Runtime 
diagnostics (Figure~\ref{fig:QPU_time}) show that QPU access time decreases with problem size and occurs 
only in discrete increments (0, 35, 70 ms), suggesting internal batching or scheduling. Remarkably, binary 
formulations such as the BLP received less QPU access time than MILP runs, despite their larger structural affinity to annealing due to the sole presence of binary variables. This observation underlines that QPU allocation is governed by internal heuristics rather 
than by problem type or size.

Balancing cost minimization with probabilistic reliability remains a central challenge. The hybrid solver 
achieved improved feasibility at higher reliability levels, crucial in power system operation, but at the 
expense of increased cost and slightly lower optimality compared to classical benchmarks. This trade-off highlights the importance of encoding design and penalty calibration in quantum optimization models, motivating systematic approaches such as the adaptive penalty factor tuning introduced here.

The QUBO reformulations revealed the primary hardware bottlenecks. In the stochastic case, the number of binary variables grows as \eqref{number}, which tends to produce more densely connected graphs that can exceed the 20-way connectivity of D-Wave’s 4{,}400-qubit Advantage2. Embedding such graphs requires long qubit chains, which amplify noise and degrade solution quality~\cite{Dwave2020}. As a result, realistic stochastic UCPs 
remain out of reach for current annealers. Even deterministic QUBO instances, though small enough to fit 
the qubit count, experience quality loss due to embedding overhead and limited coherence.

To explore this deterministic limit, we developed an adaptive penalty factor tuning algorithm that 
balances constraint enforcement and energy landscape smoothness. Applied with D-Wave’s simulated annealer, 
the method demonstrated stable convergence and provides a foundation for future quantum implementations 
once hardware connectivity improves. The simulated annealer, while classical, offered an embedding-free 
testbed to evaluate model behavior independently of quantum noise and hardware constraints.

Overall, the results suggest that QA, though not yet scalable to realistic stochastic 
problems, constitutes a promising heuristic paradigm for combinatorial optimization. Hybrid 
quantum–classical solvers already exhibit competitive behavior in large-scenario regimes, but their 
black-box design limits interpretability and reproducibility. Future progress will hinge on two fronts: 
advances in hardware, particularly higher qubit counts, improved connectivity, and reduced noise, and the 
development of transparent hybrid algorithms exposing their internal scheduling and penalty strategies.

Beyond annealing, gate-based quantum architectures might offer a more flexible framework for stochastic 
optimization. Their universality enables explicit control of subroutine design, parameter tuning, and 
error mitigation, opening the possibility of tailored stochastic optimization algorithms and systematic 
analysis of convergence and complexity. Developing such methods, especially for problems with joint chance 
constraints, represents a promising direction for future research. While this does not diminish the 
relevance of annealing for specific problem classes, gate-based approaches provide a more transparent and 
versatile platform for exploring quantum advantages in stochastic programming.

Together, these findings establish an empirical baseline for quantum and hybrid approaches to stochastic 
power system optimization. Our successive reformulations 
from MILP to QUBO delineate a conceptual bridge between classical and quantum representations of 
uncertainty. Bringing this bridge into practice will depend on future hardware advances and on the 
continued development of transparent hybrid algorithms linking probabilistic modeling and quantum 
computation within a unified optimization framework.

\section{Methods} \label{sec:methods}

This section presents the mathematical reformulations and computational procedures used to 
analyze both stochastic and deterministic variants of the UCP.  
The overall methodological workflow is summarized in Figure~\ref{fig:flow_diagram}. 
In the stochastic formulation, the demand vector is modeled as a random variable and approximated 
through scenario sampling; in the deterministic formulation, it is fixed, yielding a reduced model. 
Both follow the same transformation sequence: from a MILP, through binary encoding and constraint 
penalization, to a QUBO model suitable for quantum annealing.

\subsection{Stochastic UCP}
\label{sec:stochastic_UCP}

The stochastic formulation begins with a scenario-based approximation of the joint chance constraint 
\eqref{eq:prob}. Independent samples $\{\mathbf{D}^i\}_{i=1}^N$ represent realizations of the random 
demand. Binary indicators $y_i$ denote scenario satisfaction, leading to the constraints
\begin{align}
    \sum_{g=1}^{G} p_{g,t} &\ge D_t^i y_i, 
    && \forall i,t
    \label{eq:scenario_balance}\\
    \sum_{i=1}^{N} y_i &\ge pN, 
    \label{eq:scenario_reliability}
\end{align}
which ensure that at least a fraction $p$ of all sampled scenarios satisfies demand.  
This results in a MILP that can be solved by classical or hybrid solvers 
(e.g., Gurobi or D-Wave’s Constrained Quadratic Model solver).

Following this step, the MILP is reformulated as a Binary Linear Program (BLP) by discretizing continuous 
generation variables and representing them through binary encoding. In this representation, all continuous 
quantities—including slack variables—are integer-valued and encoded in binary form, ensuring compatibility 
with hybrid quantum–classical solvers.

To maintain linearity, both inequality constraints of the stochastic UCP are rewritten in equality form 
by introducing integer-valued slack variables. For the demand balance constraint \eqref{eq:scenario_balance}, 
we define $s_{i,t}$ such that
\begin{equation}
    \sum_{g=1}^{G} p_{g,t} = D_t^i y_i + s_{i,t}^{\text{d}}, 
    \quad s_{i,t}^{\text{d}} \in [0, s_{i,t}^{\text{d, max}}],
    \label{eq:slack_intro}
\end{equation}
with
\begin{equation}
     s_{i,t}^{\text{d, max}} = 100 \cdot \sum_{g} \!\left[ P_g^{\min} + \sum_{k=0}^{n-1} 2^k \right],
    \label{eq:slack_max}
\end{equation}
where the scaling factor $100$ enforces integer coefficients—required by D-Wave’s solvers—while preserving 
two decimal digits of accuracy. The only non-integer terms arise from the stochastic demand samples 
$D_t^i$, drawn from a multivariate normal distribution.

Analogously, the reliability constraint \eqref{eq:scenario_reliability} is written as

\begin{equation}
    \sum_{i=1}^{N} y_i = pN + s^{\text{rel}}, 
    \quad s^{\text{rel}} \in [0, s^{\text{rel, max}}],
    \label{eq:slack_reliability}
\end{equation}
where
\begin{equation}
    s^{\text{rel, max}} = N(1 - p).
\end{equation}

Then, the number of binary variables needed to encode the slack variable associated with the reliability constraint is
\begin{equation}
    N_s^{\text{reliability}} 
    = \left\lceil \log_2\!\left( N(1 - p) + 1 \right) \right\rceil.
    \label{eq:slack_reliability_max}
\end{equation}

This explicit encoding of probabilistic reliability as a deterministic binary condition makes the chance constraint directly compatible with the discrete structure required by hybrid and quantum solvers. Together, the demand and reliability slack variables define the additional binary variables required for the BLP representation.

Finally, by replacing explicit constraints with quadratic penalty terms, the BLP is transformed into a QUBO 
formulation suitable for direct execution on a quantum annealer (see Section \ref{sec:deterministic_UCP}).

\subsection{Deterministic UCP}
\label{sec:deterministic_UCP}

In the deterministic setting, demand is modeled as fixed rather than stochastic. The stochastic constraints 
\eqref{eq:scenario_balance}–\eqref{eq:scenario_reliability} are thus replaced by deterministic demand 
requirements,
\begin{equation}
    \sum_{g=1}^{G} p_{g,t} \geq D_t, \quad \forall t,
    \label{eq:det_demand_ref}
\end{equation}
where $D_t$ denotes the known demand at time period $t$ (see Appendix \ref{app:parameters_stoch_UCP}).  

To obtain a fully binary representation, continuous generation variables are discretized by binary encoding,
\begin{equation}
    p_{g,t} = P_g^{\min} u_{g,t} + \sum_{k=0}^{n-1} 2^{k} p_{g,t,k},
    \label{eq:binary_encoding_ref}
\end{equation}
where $p_{g,t,k} \in \{0,1\}$ and 
$n = \left\lceil \log_{2} \!\left( \max_{g} \{ P_g^{\max} - P_g^{\min} \} + 1 \right) \right\rceil$ 
is the number of bits required to span the admissible generation range. Note that this choice of $n$ allows $p_{g,t}$ to reach values larger than $P_g^{\max}$, but this is prevented by the capacity constraint in Equation \eqref{eq:capacity}.

Using a binary encoding requires the addition of an extra constraint that couples the status variables $u_{g,t}$ with the variables $p_{g,t,k}$. This is, imposing that when a generator is in operation, at least one variable $p_{g,t,k}$ has to be 1,

\begin{equation}
    \sum_{g=1}^{G} \sum_{t=1}^{T} \left[ (1 - u_{g,t})\sum_k p_{g,t,k} \right].
\end{equation}

All remaining operational and logical constraints are then incorporated into the objective function as 
quadratic penalties, leading to a QUBO formulation.
Equality constraints are enforced by penalizing squared deviations,
\begin{equation}
    \lambda \left( u_{g,t} - u_{g,t-1} - z_{g,t}^{\text{on}} + z_{g,t}^{\text{off}} \right)^{2},
\end{equation}
while inequalities are expressed through binary-encoded slack variables, for example:
\begin{align}
    \sum_{g=1}^{G}\left[ P_g^{\min} u_{g,t} + \sum_{k=0}^{n-1} 2^{k} p_{g,t,k} \right] - D_t - \sum_{k} 2^k s_{t,k} &= 0, 
    \quad s_{t,k} \in \{0,1\}, \\
    \lambda \left( \sum_{g=1}^{G} \left[ P_g^{\min} u_{g,t} + \sum_{k=0}^{n-1} 2^{k} p_{g,t,k} \right] - D_t - \sum_{k} 2^k s_{t,k} \right)^{2}.
\end{align}

All logical, capacity, and ramping constraints are treated analogously.
Each inequality constraint—demand balance, ramping limits, and capacity bounds—is represented 
in equality form using binary-encoded slack variables. The number of bits required for each slack 
is determined by the admissible range of the constraint; detailed derivations are provided in 
Appendices \ref{app:slack_number}, \ref{app:slack_det_demand}, \ref{app:slack_ramp}, \ref{app:slack_capacity}.

This procedure converts the deterministic UCP into an unconstrained quadratic program,
\begin{equation}
\begin{aligned}
    \min_{\mathbf{x}} \quad &
    \sum_{g=1}^{G} \sum_{t=1}^{T}
    \Big[
        z_{g,t}^{\text{on}} C_{g}^{\text{StartUp}}
        + z_{g,t}^{\text{off}} C_{g}^{\text{ShutDown}}
        + C_g u_{g,t}
        + b_g ( P_{g}^{\min} u_{g,t} + \sum_{k=0}^{n-1} 2^{k} p_{g,t,k} )
    \Big] \\
    &+ \sum_{i} \lambda_i (\text{constraint violation})^{2},
\end{aligned}
\label{eq:deterministic_QUBO}
\end{equation}
where each penalty term corresponds to a constraint expressed in binary form.  
The resulting QUBO can be solved using either simulated annealing or quantum annealing, providing 
a deterministic benchmark for the stochastic formulations.

\subsection{Adaptive penalty factor tuning}
\label{methods:lagrange_tuning}

In the QUBO reformulation of both stochastic and deterministic UCPs, all operational and logical 
constraints are enforced through quadratic penalty terms. The corresponding penalty factors determine the relative weight between the objective and constraint satisfaction. 
Their choice is critical: excessively small factors produce infeasible solutions, whereas overly large values distort the objective landscape, creating steep, fragmented regions that hinder convergence and increase the likelihood of local trapping. A systematic and adaptive procedure for tuning these factors 
is therefore essential to balance feasibility and optimality.

\medskip
Let $C$ denote the total number of constraints and $\mathcal{N}$ the number of solutions obtained in one solver call 
(sample set). For each constraint $i = 1, \dots, C$, we define the \emph{feasibility ratio}
\begin{equation}
    R_i = \frac{n_i}{\mathcal{N}},
    \label{eq:feasibility_ratio}
\end{equation}
where $n_i$ is the number of feasible solutions satisfying constraint $i$. This ratio provides an empirical 
measure of the proportion of constraint satisfaction across the sampled population.

To ensure overall feasibility, we define a global stopping threshold
\begin{equation}
    K = \frac{C - 1}{C} + \frac{1}{\mathcal{N}},
    \label{eq:feasibility_threshold}
\end{equation}
which guarantees that at least one constraint is satisfied across the sample set.
If $R_i \ge K$ for all $i$, at least one feasible configuration exists in the ensemble.

\medskip
The tuning process begins with all factors initialized uniformly as
\[
\lambda_i^{(0)} = 1, \qquad i = 1,\dots,C.
\]
At each iteration, the solver is executed using the current set of factors, and a new population of solutions is sampled. Based on these outcomes, feasibility ratios $R_i$ are computed and factors updated according to the rule
\begin{equation}
    \lambda_i^{(k+1)} = 
    \lambda_i^{(k)} \bigl( 1 + S(A, \kappa, R_0, R_i^{(k)}) \bigr),
    \label{eq:lambda_update}
\end{equation}
where the update factor $S$ is a sigmoidal function,
\begin{equation}
    S(A, \kappa, R_0, R_i) = \frac{A}{1 + e^{\kappa (R_i - R_0)}}.
    \label{eq:sigmoid}
\end{equation}
The function parameters $(A, \kappa, R_0)$ control the amplitude, steepness, and centering of the 
adjustment, respectively. Large positive $\kappa$ values yield a sharper transition between weak and strong 
updates, while $R_0$ defines the reference feasibility level at which the correction strength begins to 
decrease. The choice of $A$ determines the maximum multiplicative increment applied to the factor. 

\medskip
The sigmoidal structure ensures smooth, bounded updates: when the feasibility ratio $R_i$ falls well below 
the reference $R_0$, the corresponding factor $\lambda_i$ increases sharply, penalizing constraint 
violations more strongly; when $R_i$ approaches or exceeds $R_0$, the adjustment becomes mild, thereby 
avoiding numerical instability or over-penalization. In this manner, the update mechanism acts as an 
adaptive feedback control on constraint satisfaction.

\medskip
The iterative tuning proceeds until either all constraints reach the global feasibility threshold 
\eqref{eq:feasibility_threshold}, or a predefined maximum number of iterations is attained. The resulting set 
of factors $\{\lambda_i^\ast\}$ reflects the empirical difficulty of satisfying each constraint within 
the solution space: tighter or more frequently violated constraints are assigned higher weights, while 
those that are naturally satisfied remain weakly penalized.

\medskip
This adaptive procedure provides two main benefits. First, it automates the calibration of penalty weights 
without manual intervention, ensuring that each constraint receives an appropriate penalization level 
commensurate with its violation frequency. Second, it improves the convergence properties of both simulated 
and quantum annealing by shaping a balanced energy landscape: feasible regions remain reachable, while 
infeasible minima are progressively suppressed. The resulting tuned QUBO formulation thus constitutes a 
well-scaled, hardware-compatible representation of the original optimization problem.

\section*{Data availability Statement}
The data used to define the models was originally adopted from \cite[Ch.~7]{Conejo}. Data regarding scripts and results analyzed in this work can be found in the Github repository \cite{Ribes2025}.

\def\urlprefix{}\def\href#1#2#3#4{\ifstrequal{#2}{[link]}{}{#2\newline}}
\bibliographystyle{naturemag}
\bibliography{refs}

\section*{Author contributions}
T.G.G. conceived the main idea and supervised the work. D.M. developed and implemented the code on D-Wave, performed benchmarking studies. All authors analysed the results and contributed to the writing and review of the manuscript.

\section*{Funding}
This research received no external funding.

\section*{Competing Interests}
The authors declare that they have no competing financial or non-financial interests. 

\newpage
\section*{Appendices}
\appendix

\section{Model Parameters}
\label{app:parameters_stoch_UCP}

This appendix reports the numerical data used for the stochastic and deterministic UCP instances.

\begin{table}[H]
    \centering
    
    \begin{tabular}{|l|c|c|c|}
        \hline
        \textbf{Unit \#} & \textbf{1} & \textbf{2} & \textbf{3} \\
        \hline
        $P_g^{\min}$ [MW] & 50  & 80  & 40  \\[0.1cm]
        $P_g^{\max}$ [MW] & 350 & 200 & 140 \\[0.1cm]
        $R^{\text{down}}_{g}$ [MW/h] & 300 & 150 & 100 \\[0.1cm]
        $R^{\text{up}}_{g}$ [MW/h]   & 200 & 100 & 100 \\[0.1cm]
        $C^{\text{StartUp}}_{g}$ [\$] & 20  & 18  & 5   \\[0.1cm]
        $C^{\text{ShutDown}}_{g}$ [\$] & 0.5 & 0.3 & 1.0 \\[0.1cm]
        $b_g$ [\$/MWh] & 0.10 & 0.125 & 0.150 \\[0.1cm]
        $C_g$ [\$]     & 5 & 7 & 6 \\[0.1cm]
        \hline
    \end{tabular}
    \caption{Generator characteristics.}
    \label{tab:miqpparameters_values_stochastic}
\end{table}

\begin{table}[H]
    \centering
    \begin{tabular}{|c|c|c|c|}
        \hline
        \textbf{Time period} & \textbf{1} & \textbf{2} & \textbf{3} \\ \hline
        $\mu_t$ [MW]         & 225 & 630 & 400 \\ \hline
        $\sigma_t$ [MW]      & 25  & 40  & 28  \\ \hline
    \end{tabular}
    \caption{Parameters of the multivariate normal demand distribution.}
    \label{tab:demands_stochastic}
\end{table}

\begin{table}[H]
    \centering
    \begin{tabular}{|c|c|c|c|}
        \hline
        \textbf{Unit \#} & \textbf{1} & \textbf{2} & \textbf{3} \\ \hline
        $u_{g,0}$ & 0 & 0 & 1 \\ \hline
        $p_{g,0}$ [MW] & 0 & 0 & 100 \\ \hline
    \end{tabular}
    \caption{Initial on/off and output states of the generating units.}
    \label{tab:priors_stochastic}
\end{table}

\begin{table}[H]
    \centering
    \begin{tabular}{|c|c|c|c|}
        \hline
        \textbf{Time period} & \textbf{1} & \textbf{2} & \textbf{3} \\ \hline
        $D_t$ [MW] & 160 & 500 & 400 \\ \hline
    \end{tabular}
    \caption{Deterministic demand profile.}
    \label{tab:demands_deterministic}
\end{table}

\section{Computation of Binary Slack Variables}
\label{app:slack_number}

In the binary formulations, every inequality constraint is expressed in equality form by introducing an integer slack variable. The number of binary variables required to represent a given slack depends on the admissible range of its integer values.

Let $s^{\max}$ denote the largest integer value that the slack can attain without violating the constraint (i.e., when the corresponding penalty term vanishes). A binary representation requires $n_s$ bits such that
\begin{equation}
    2^{n_s-1} \le s^{\max} < 2^{n_s}.
\end{equation}
Hence, the minimal sufficient number of bits is
\begin{equation}
    n_s = \left\lceil \log_2(s^{\max} + 1) \right\rceil.
    \label{eq:ns_general}
\end{equation}
The following subsections summarize the derivations of $s^{\max}$ and $n_s$ for the main constraint classes.

\subsection{Deterministic UCP: Demand Constraint}
\label{app:slack_det_demand}

The deterministic demand balance
\begin{equation}
    \sum_{g=1}^{G} p_{g,t} \ge D_t, \quad t = 1,\dots,T,
\end{equation}
is rewritten as
\begin{equation}
    \sum_{g=1}^{G} p_{g,t} = D_t + s_t, \qquad s_t \in [0, s^{\max}],
\end{equation}
where the maximum slack corresponds to all units producing their maximum binary-encoded output,
\begin{equation}
    s^{\max} = \sum_g \left[ P_g^{\min} + \sum_{k=0}^{n-1} 2^k \right].
\end{equation}
Because $D_t$ is deterministic and integer-valued, no scaling factor is required. The total number of binary variables needed to encode these slacks is
\begin{equation}
    N_s^{\text{demand}} = T \left\lceil 
        \log_2 \!\left(
            \sum_g \!\left[ P_g^{\min} + \sum_{k=0}^{n-1} 2^k \right] + 1
        \right) 
    \right\rceil.
\end{equation}

\subsection{Ramping Limits}\label{app:slack_ramp}

The ramping constraints
\begin{equation}
    -R^{\text{down}}_g \le p_{g,t} - p_{g,t-1} \le R^{\text{up}}_g,
    \qquad g = 1,\dots,G,\ t = 1,\dots,T,
\end{equation}
are reformulated as
\begin{equation}
    p_{g,t} - p_{g,t-1} = s_{g,t}, \quad 
    s_{g,t} \in [-R^{\text{down}}_g, R^{\text{up}}_g].
\end{equation}
The number of bits required for each ramping slack variable follows directly from
\begin{equation}
    n_s^{g} = \left\lceil \log_2(R^{\text{up}}_g + R^{\text{down}}_g + 1) \right\rceil,
\end{equation}
and the total number of binary variables associated with ramping constraints is
\begin{equation}
    N_s^{\text{ramp}} = 
    T \sum_g n_s^{\text{ramp},g}
    = T \sum_g \left\lceil 
        \log_2(R^{\text{up}}_g + R^{\text{down}}_g + 1) 
    \right\rceil.
\end{equation}

\subsection{Capacity Limits}
\label{app:slack_capacity}

The generator capacity limits
\begin{equation}
    P_g^{\min} u_{g,t} \le p_{g,t} \le P_g^{\max} u_{g,t}
\end{equation}
ensure that production remains within feasible operating bounds. Using the binary encoding of $p_{g,t}$ and the coupling constraint, the inequality can be reduced to
\begin{equation}
    p_{g,t} \le P_g^{\max}.
\end{equation}
Introducing a non-negative slack variable $s_{g,t}$, the equality form becomes
\begin{equation}
    p_{g,t} + s_{g,t} = P_g^{\max}, 
    \quad s_{g,t} \in [0, s_g^{\max}],
\end{equation}
with $s_g^{\max} = P_g^{\max}$. The number of binary variables required for this representation is
\begin{equation}
    n_s^{\text{capacity},g} = 
        \left\lceil \log_2(P_g^{\max} + 1) \right\rceil,
\end{equation}
and the total number of capacity-related binary variables is
\begin{equation}
    N_s^{\text{capacity}} = 
    T \sum_g n_s^{\text{capacity},g} 
    = T \sum_g \left\lceil 
        \log_2(P_g^{\max} + 1) 
    \right\rceil.
\end{equation}

\end{document}